\documentclass[journal]{IEEEtran}

\usepackage{graphicx}
\usepackage{epstopdf}
\usepackage{amsmath}
\usepackage{amssymb}
\usepackage{amsthm }
\usepackage{epsfig}
\usepackage{times}
\usepackage{setspace}
\usepackage{bm}
\usepackage{siunitx}
\usepackage{threeparttable}
\usepackage{balance}
\usepackage{tabularx}
\usepackage{subcaption}
\usepackage{balance}

\usepackage{array}

\usepackage[T1]{fontenc}
\usepackage{times}
\usepackage[mathscr]{eucal}
\usepackage{wrapfig}
\usepackage{multirow}
\usepackage{tablefootnote}

\newcommand{\equals}{\!=\!}
\newcommand{\lteq}{\!\le\!}
\newcommand{\minus}{\!-\!}

\newcommand{\gthan}{\!>\!}
\newcommand{\lthan}{\!<\!}
\newcommand{\plus}{\!+\!}

\newcommand{\Tsym}{T_\mathrm{sym}}
\begin{document}

\title{The LoRa Modulation Over Rapidly-Varying Channels: Are the Higher Spreading Factors \\ Necessarily More Robust? }

\author{Harishwar Reddy Bapathu\IEEEauthorrefmark{1} and Siddhartha S. Borkotoky\IEEEauthorrefmark{2}\\
\IEEEauthorrefmark{1}National Institute of Technology Tiruchirappalli, India\\
\IEEEauthorrefmark{2}Indian Institute of Technology Bhubaneswar, India\\
\texttt{harishwar2512@gmail.com, borkotoky@iitbbs.ac.in} \vspace*{-4mm}
}

\maketitle

\begin{abstract}
The chirp spread spectrum (CSS) modulation scheme is employed by the physical layer of the Long Range (LoRa) communication technology. In this paper, we examine the performance of CSS  over time-varying channels whose gain may change during the reception of a LoRa frame. This is in contrast to the usually employed model in the literature, which assumes the channel gain to be constant throughout a frame. Specifically, we investigate the effects of exponentially correlated Rayleigh fading on the frame-error rate of a CSS receiver in which the channel gain is estimated at the beginning of each frame. Our primary observation is that over rapidly-varying channels, the robustness benefits of the larger spreading factors tend to disappear as the payload size grows. This observation, which is contrary to the common perception that higher spreading factors necessarily provide greater immunity against  noise, highlights the need to consider channel characteristics and payload sizes in allocating the spreading factor for reliable and energy-efficient LoRa communications.          
\end{abstract}

\section{Introduction}
The LoRa communication technology has received significant attention in the recent past as a key enabler of the Internet of Things. It has found applications in the domains of smart city~\cite{PBF18}, smart metering~\cite{CSG18}, agriculture~\cite{JYC19}, and Industry 4.0~\cite{SGI20}, to name a few. One of the major factors driving LoRa's widespread adoption is its capability to provide robust communications over long ranges at low power levels. These features are attributable to LoRa's PHY layer, which employs CSS, a spread-spectrum modulation technique that possesses several desirable characteristics such as high resistance to Doppler and multipath and low receiver sensitivity~\cite{Sem15}.    

The growing interest in LoRa has led to a large body of experimental and theoretical work on its frame-delivery performance. Substantial attention has been devoted to analyzing the effects of interference in LoRa networks, both from a link-level (e.g.,~\cite{BRV16}--\cite{MSG19}) and PHY-level~(e.g., \cite{DNB19}) perspective. By contrast, the effects of channel artifacts -- such as the temporal variations in the channel gain introduced by fading -- have been relatively unexplored. Indeed, many works on interference analysis take into account the effects of fading, but a block-fading model is typically assumed. In these block-fading models, the channel gain is assumed to be constant throughout the reception of a LoRa frame but can change from  one frame to the next. This is a reasonable assumption for short frame lengths and slowly varying channels. However, owing to the low bit rate of LoRa, a LoRa frame can be quite long depending on the payload size. For example, the duration of LoRa frame carrying a payload of 50 bytes and using a spreading factor 12 and occupying a bandwidth of 125 KHz is in excess of 1.6 seconds. 
For a long frame  transmitted over a highly dynamic channel, variations in the channel gain during the frame's reception cannot be ruled out. Because a typical receiver estimates the channel gain using pilot symbols included at the beginning of the frame~\cite{Gol05}, such variations may result in a mismatch between the actual and estimated gain, thus increasing the probability of frame error.   

In LoRa, a parameter called the spreading factor is typically varied in order to obtain a desired robustness against noise and fading.
The spreading factor is an integer between 7 and 12~\cite{Sem15}. The higher the spreading factor, the more robust the communication is assumed to be, since higher spreading factors lead to lower receiver sensitivity~\cite{GeR17}. Therefore, spreading factor 12 is often the default setting for LoRa~\cite{Sem19}.  The performance  benefit of larger spreading factors comes at the cost of a longer on-air time, since for a fixed payload size, increasing the spreading factor by 1 doubles the LoRa symbol duration. Therefore, in a fast-fading environment, a larger spreading factor makes it more likely that the channel gain might change within the duration of the frame, thereby making the receiver's estimate of the channel gain inaccurate. This has the potential to degrade the demodulator's performance and negate the advantages of the higher spreading factors.    

Motivated by this reasoning, we perform a simulation study of CSS over fading channels whose gain can change within a frame. We investigate DFT-based CSS demodulation~\cite{DNB19} over a Rayleigh-fading channel whose gain can change from sample to sample within a frame in an exponentially correlated manner. Our main observation is that under certain fading conditions and for some payload sizes, the larger spreading factors suffer from higher frame-error rate than the smaller ones, which indicates that simply increasing the spreading factor to improve the probability of frame delivery may not always be the optimal strategy.   

The rest of the paper is organized as follows: We provide a brief description of CSS modulation and demodulation in Section~\ref{CSS_description}, followed by a description of the simulation model in~\ref{fading_model}. Numerical results are presented in Section~\ref{sim_results}, and Section~\ref{conclusion} concludes the paper.   

\section{CSS Modulation and Demodulation} \label{CSS_description}
We summarize the important features of CSS in this section. More detailed discussions can be found, for example, in~\cite{DNB19} and~\cite{ChE19}. 

\vspace*{-4mm}
\subsection{CSS modulation}
The CSS signal set is an $M$-ary orthogonal signal set characterized by a \textit{spreading factor} $S$. A CSS symbol, also referred to as a \textit{chirp}, represents $S$ bits and has a duration 
\begin{equation}
\Tsym(S) = 2^S/W,
\end{equation}
where $W$ is the transmission bandwidth. A chirp is a constant-envelop sinusoidal signal whose frequency sweeps linearly through the bandwidth $W$ over the duration  $\Tsym(S)$. Let $f_c$ denote the center frequency of the LoRa waveform. The chirp corresponding to the all-zeros bit pattern (also referred to as symbol 0 or the \textit{basic chirp}) begins with an instantaneous frequency $f_c \minus W/2$ and increases the frequency  linearly to reach a maximum value of $f_c \plus W/2$ at the end of the chirp. The baseband equivalent of the basic chirp can be expressed as~\cite{DNB19}
\begin{equation}
    x_0(t) = \exp\left\{j2\pi\left( \frac{\mu t}{2} - \frac{W}{2} \right) t\right\}, \quad t\in[0,\Tsym(S)],
\end{equation}
where $\mu \equals W/\Tsym(S)$ is the \textit{chirp-rate}. Since each symbol carries $S$ bits, there is a total of $M \equals 2^S$ unique CSS symbols. 
Symbols 1 through $M \minus 1$ are cyclically shifted versions of symbol 0. Specifically, symbol $m$ is obtained by cyclically shifting symbol 0 in time by an amount $m/W$, i.e.,
\begin{equation}
    x_m(t) = x_0(\mathrm{mod}(t-\frac{m}{W},\Tsym(S))),\quad t\in[0,\Tsym(S)],
\end{equation}
for $m = 0,1,\ldots,M \minus 1$.

\vspace*{-2mm}
\subsection{CSS demodulation}
\subsubsection{Demodulation over an AWGN Channel}
We first summarize the procedure for the DFT-based demodulation of a CSS symbol over an additive white Gaussian noise (AWGN) channel. A detailed description can be found in~\cite{DNB19}. 

In DFT-based demodulation, the received signal is first sampled at a rate \mbox{$f_s= 1/W$}. For CSS symbol $m$ (i.e., $x_m(t)$) received over an AWGN channel, the sampled values  are
\begin{equation}
    r[n] = x_m[n] + w[n], \quad n = 0, \ldots, M-1,
\end{equation}
where $x_m[n]$ are the sampled values of  $x_m(t)$, and $w[n]$ are i.i.d. complex Gaussian noise samples with mean 0 and standard deviation $\sigma$. The samples $x_m[n]$ can be expressed as $x_m[n] = x_0[\mathrm{mod} (n \minus m,M)]$, where $x_0[n]$ are the sampled values of the basic baseband chirp and are given by~\cite{DNB19}
\begin{equation}
    x_0[n] = \exp \left\{j2\pi\left( \frac{n^2}{2M} - \frac{n}{2}\right)\right\}; \quad n = 0,1,\ldots,M-1. 
 \end{equation}
The receiver obtains the sequence $y[n]$ by multiplying $r[n]$ with the conjugate of $x_0[n]$, that is
\begin{equation}
    y[n] = r[n]x_0^*[n]; \quad n = 0,1,\ldots,M-1. 
\end{equation}
Next, an $M$-point DFT of $y[n]$ is taken. Let $Y[k]$, \mbox{$k \equals 0,1,\ldots M \minus 1$}, denote the DFT output. The receiver decides symbol $\hat m$ was sent if
\begin{equation}
    \hat m = \mathrm{arg} \{ \max_{0 \leq k \leq M-1} |Y[k]|\}.
\end{equation}

\subsubsection{Demodulation over a Fading Channel}
The sampled values of the received signal over a fading channel are
\begin{equation}
    r[n] = h[n]x_m[n] + w[n], \quad n = 0, \ldots, M-1,
\end{equation} 
where $h[n]$ is a complex fading coefficient. We employ the common detection procedure over a fading channel, in which each sample $r[n]$ is multiplied by the conjugate of $\hat h[n]$, which is the the receiver's estimate of the fading gain~\cite{Gol05}. That is,
\begin{equation}
    \tilde r[n] = \hat h^*[n] r[n]. 
\end{equation}
The remainder of the procedure is the same as that for the AWGN channel. The receiver obtains $Y[k]$ by taking the $M$-point DFT of the sequence $y[n] = \tilde r[n]*x_0[n]$ and declares symbol $\hat m$ to be the transmitted symbol if the $\hat m$-th sample of $Y[k]$ is the largest of all $M$ samples.

\section{Simulation Setup} \label{fading_model}
We simulate the reception of a sequence of CSS modulated \textit{frames} over a fading channel. Each frame carries $B$ bytes of CSS-modulated data. Thus, a frame with spreading factor $S$ carries \mbox{$8B/S$} CSS symbols. A transmission bandwidth of 125 KHz, which is typical for most LoRa implementations, is employed. It follows that the sampling interval for the demodulator is 8 microseconds. A frame is treated as received correctly if each CSS symbol in the frame is correctly detected. Our performance measure is the \textit{frame-error rate} (FER), which is the fraction of the frames received in error. For each data point in our FER curves, 50,000 frames are simulated.    

Without loss of generality, each CSS symbol is assumed to be transmitted with unit power. We define the signal-to-noise ratio ($\mathrm{SNR}$) as the ratio of the transmit power per bit to the noise variance. Recall that each CSS symbol carries $S$ bits and the noise variance is $\sigma^2$. Thus, as per our definition, $\mathrm{SNR} \equals 1/S\sigma^2$, which is the same as the expression in~\cite{DNB19}. 
\vspace{-2mm}
\subsection{Exponentially-Correlated Rayleigh Fading}
We simulate an exponentially correlated Rayleigh-fading channel~\cite{HuN09}--\cite{KaS94}, for which the fading coefficient $h[n]$ is a complex Gaussian random variable with mean 0 and variance 1. The fading is independent from one frame to the next. However, the samples within a frame are correlated. The autocorrelation between the $n$-th and the $m$-th sample in the frame is equal to
\begin{equation}
    \mathrm{Cov}\{h[m],h[n]\} = q^{|m-n|},
\end{equation}
where $0 \leq q  \leq 1$ is the \textit{covariance parameter} of the channel. Note that the autocorrelation and the autocovariance are the same due to the zero mean of $h[n]$. For $q \equals 0$, we obtain a channel whose gain varies independently from sample to sample, whereas for $q \equals 1$, we obtain a block-fading channel whose gain is constant throughout the frame. In general, the smaller the value of $q$, the more rapidly the channel gain fluctuates within a frame. Our primary interest is in values of $q$ very close to 1, so that consecutive samples are highly correlated. This is because, even for the most rapidly varying channel, we do not expect the channel gain to change  by much over the sampling interval of 8 microseconds. Instead, we expect gradual variations in the channel gain during the frame's reception. Consequently, we restrict attention to $0.9999 \lteq q \lteq 1$ in our simulations.  

We assume that the receiver estimates the fading gain at the beginning of each frame, as is typically done in most packet-based wireless communication systems~\cite{Gol05}. We also make the simplifying assumption that the receiver has a perfect estimate of the fading at the beginning of the frame. Thus, for $q \equals 1$, the receiver has exact knowledge of the channel gain for every sample in the frame. For $q \lthan 1$, the receiver starts out with an accurate estimate of the channel gain, but the estimate may become inaccurate for later samples in the frame.

\begin{figure} 
    \centering
    \includegraphics[scale=0.27, bb=-50 30 820 570]{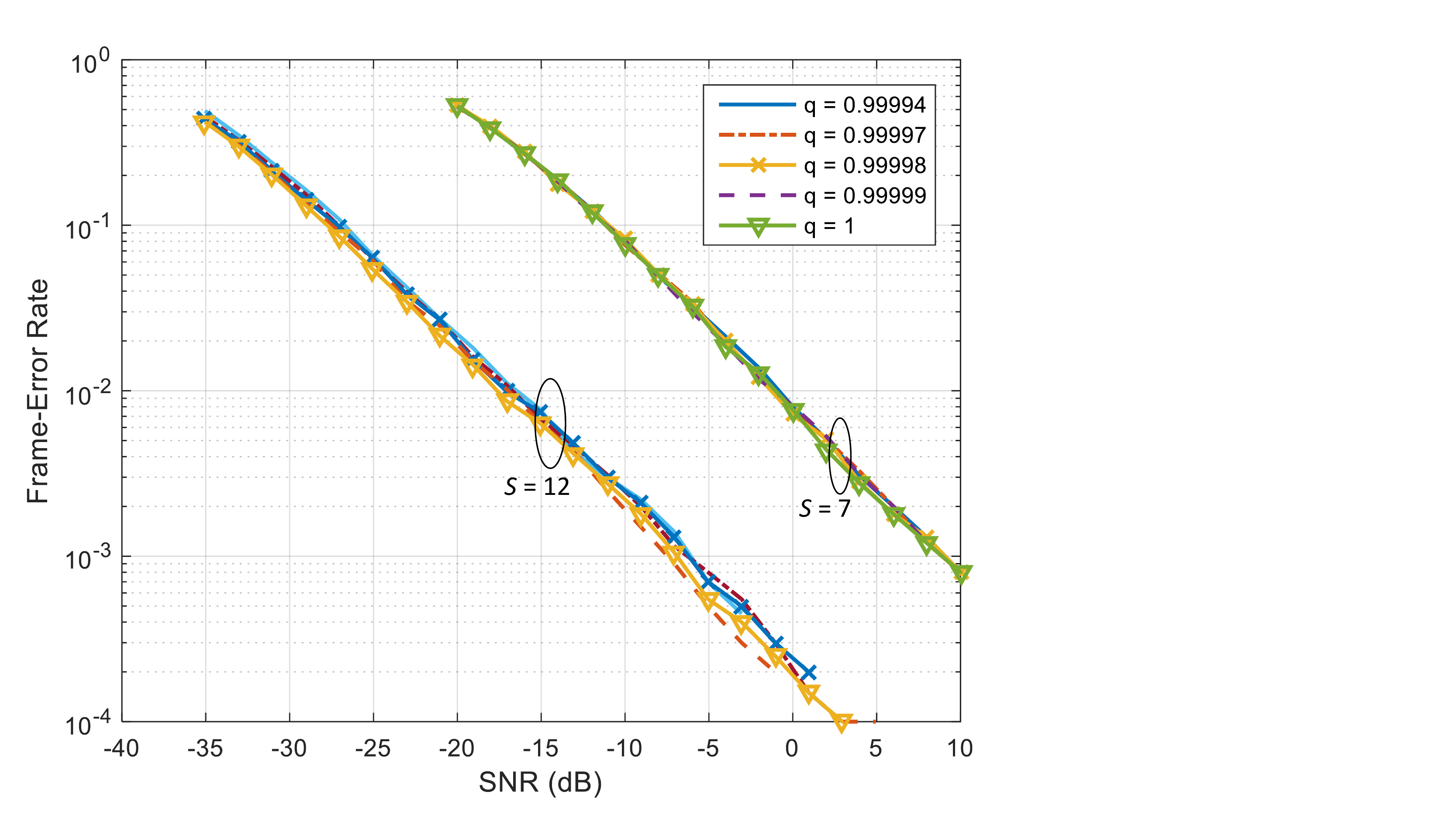}
    \setlength{\belowcaptionskip}{-12pt}
    \caption{FER comparison for $B \equals 1$.}
    \label{Fig1}
\end{figure}

\begin{figure}
    \centering
    \includegraphics[scale=0.27, bb=-50 40 820 585]{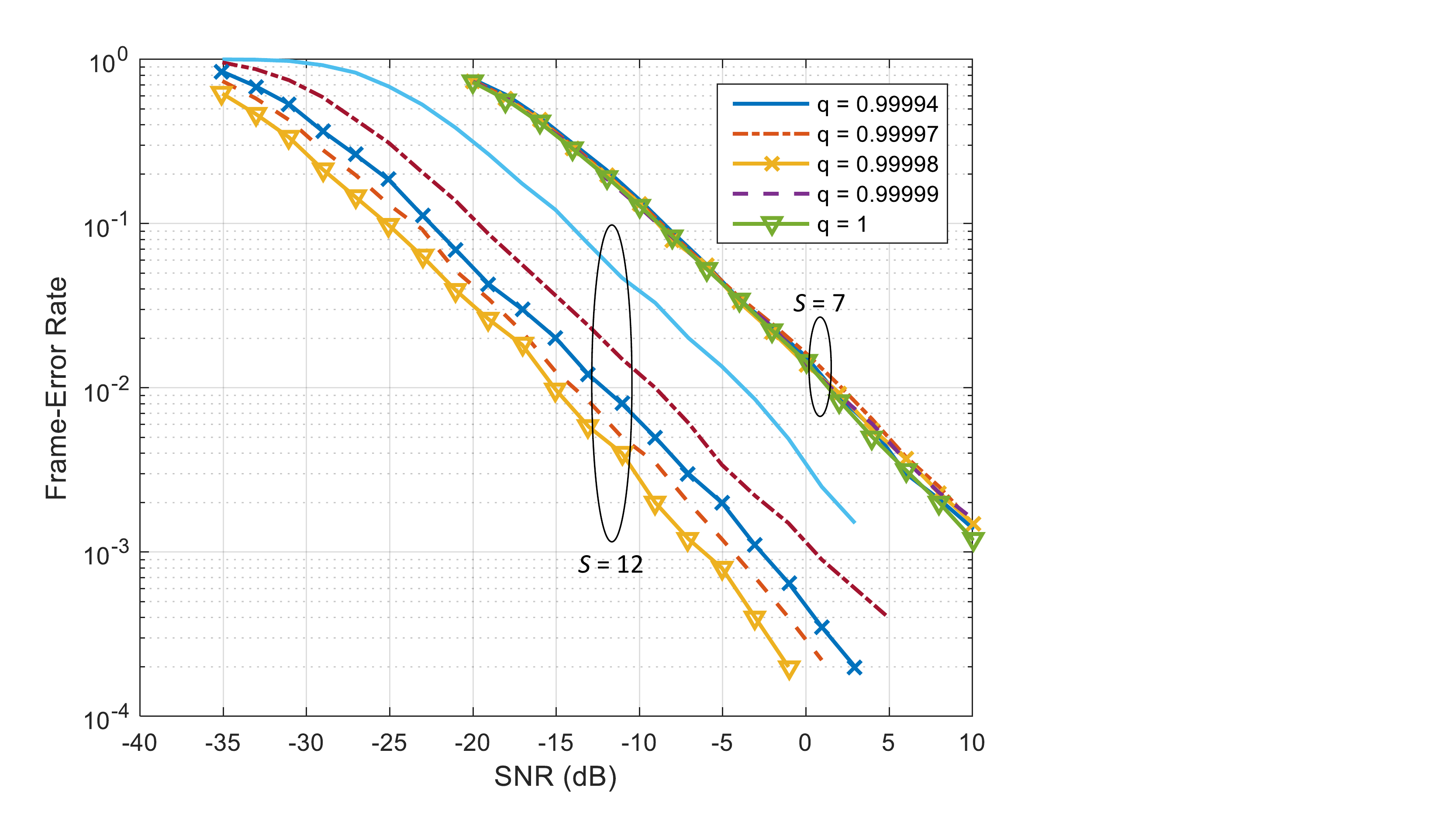}
    \setlength{\belowcaptionskip}{-10pt}
    \caption{FER comparison for $B \equals 10$.}
     \label{Fig2}
\end{figure}

\begin{figure} 
    \centering
    \includegraphics[scale=0.27, bb=-50 30 820 570]{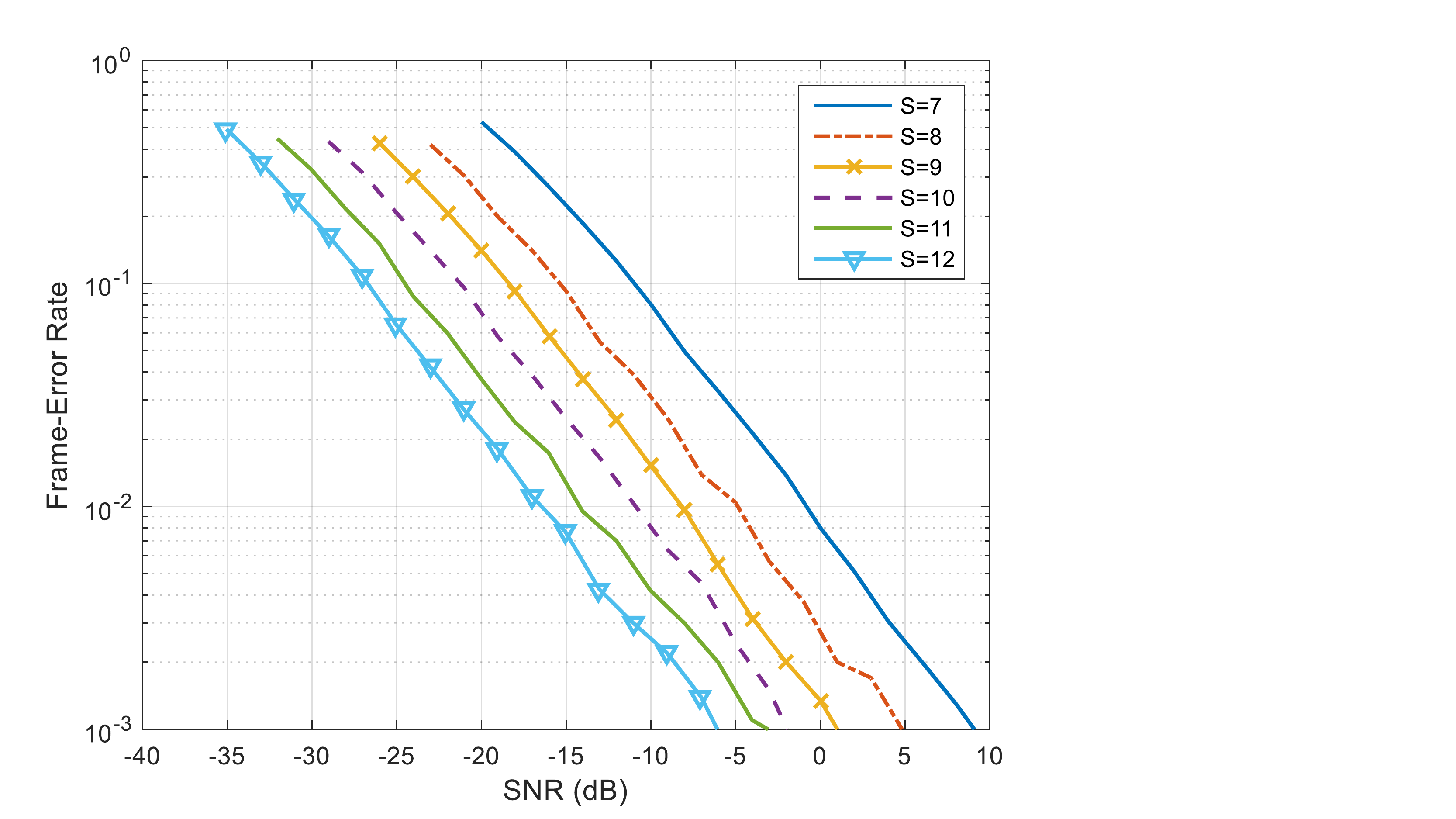}
    \setlength{\belowcaptionskip}{-12pt}
    \caption{FER comparison for $q \equals 0.99994$ and $B \equals 1$.}
    \label{Fig3}
\end{figure}

\begin{figure} 
    \centering
    \includegraphics[scale=0.27, bb=-50 30 820 570]{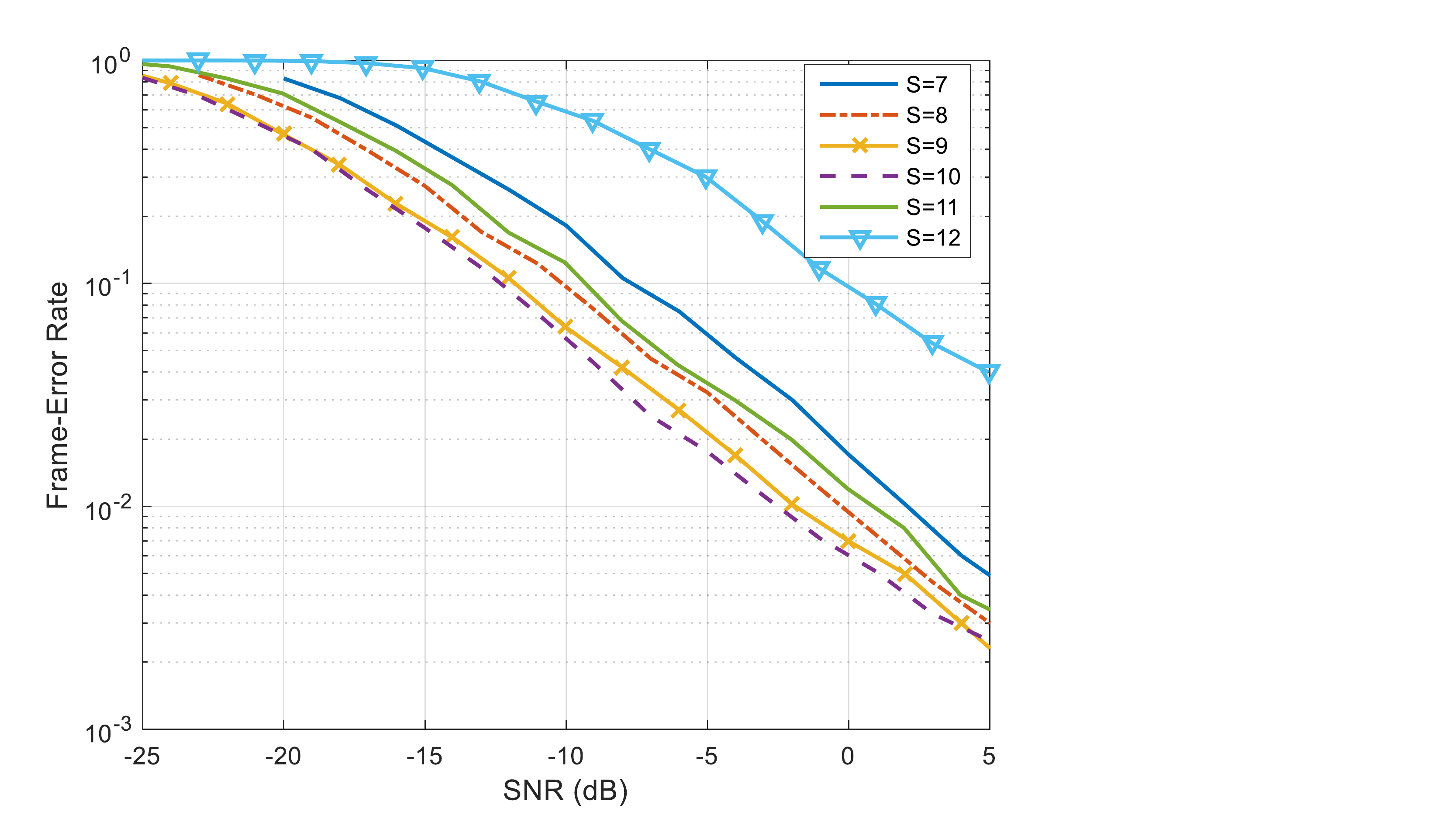}
    \setlength{\belowcaptionskip}{-12pt}
    \caption{FER comparison for $q \equals 0.99994$ and $B \equals 20$.}
    \label{Fig4}
\end{figure}

\vspace{-0mm}
\section{Numerical Results} \label{sim_results}
Fig.~\ref{Fig1} shows the FER for two spreading factors $S\equals 7$ and $S \equals 12$, which are the smallest and largest spreading factors, respectively, used by LoRa. Each frame carries a payload of $B \equals 1$ byte. FER-vs.-SNR curves are plotted for five different values of the covariance parameter $q$. Recall that $q\equals1$ yields a channel whose gain is constant throughout the frame, whereas a smaller $q$ results in exponentially correlated random variations in the channel gain within the frame. We observe that for a given spreading factor, the FER performance doesn't change much with a change in $q$. We also observe that for each value of $q$, $S\equals 12$ provides lower FER than $S=7$, which is consistent with the common belief that the larger spreading factors facilitate more robust communications.

In Fig.~\ref{Fig2}, we plot the same curves as in Fig.~\ref{Fig1}, but for a payload size of $B \equals 10$ bytes. As before, the FER for $S \equals 7$ does not change significantly with a change in $q$; but for $S \equals 12$, we observe drastic performance degradations as $q$ is reduced. There is an order-of-magnitude increase in the FER as $q$ is reduced from 1 to 0.99994. Recall that for the same payload size, a frame with $S \equals 12$ is much longer than  a frame with $S \equals 7$. Thus, there is a greater chance that the channel gain towards the end of the frame with $S \equals 12$ will be substantially different from the receiver's estimate of the gain, which was obtained at the beginning of the frame.

The FER curves for all spreading factors between $7$ and $12$ for a payload size of $B \equals 1$ byte and a covariance parameter $q \equals 0.99994$ are shown in Fig.~\ref{Fig3}. For this scenario, the larger spreading factors consistently provides lower FER than the smaller spreading factors. In Fig.~\ref{Fig4}, we plot the same curves for a payload size of $B \equals 20$ bytes. In this case, the performance pattern is quite different. The FER decreases as the spreading factor is increased from $7$ through $10$, but that trend is broken as we increase the spreading factor further. The FER for $S \equals 11$ is not lower than that for $S \equals 10$; instead, it is between $S \equals 7$ and $S \equals 8$. For $S \equals 12$, the FER is worse than that for all other spreading factors. Thus, in this situation, the extra energy spent by using $S \equals 12$ does not provide any performance benefits over other spreading factors; rather, the performance is actually worse. 

\begin{figure} 
    \centering
    \includegraphics[scale=0.28, bb=-50 30 820 570]{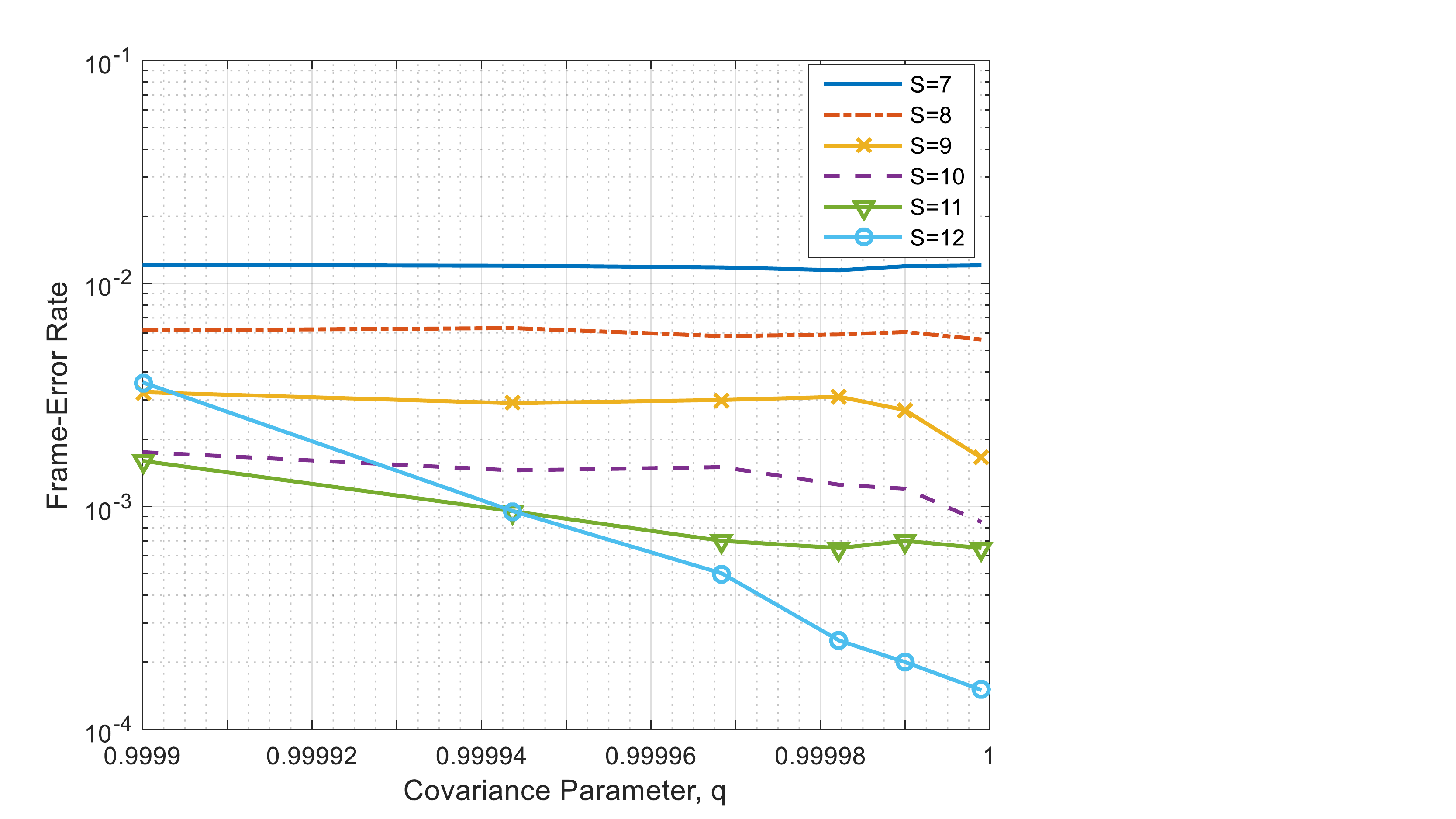}
    \setlength{\belowcaptionskip}{-11pt}
    \caption{FER as a function of $q$ (\mbox{$B \equals 5$.)}}
    \label{Fig5}
\end{figure}

\begin{figure} 
    \centering
    \includegraphics[scale=0.28, bb=-50 30 820 570]{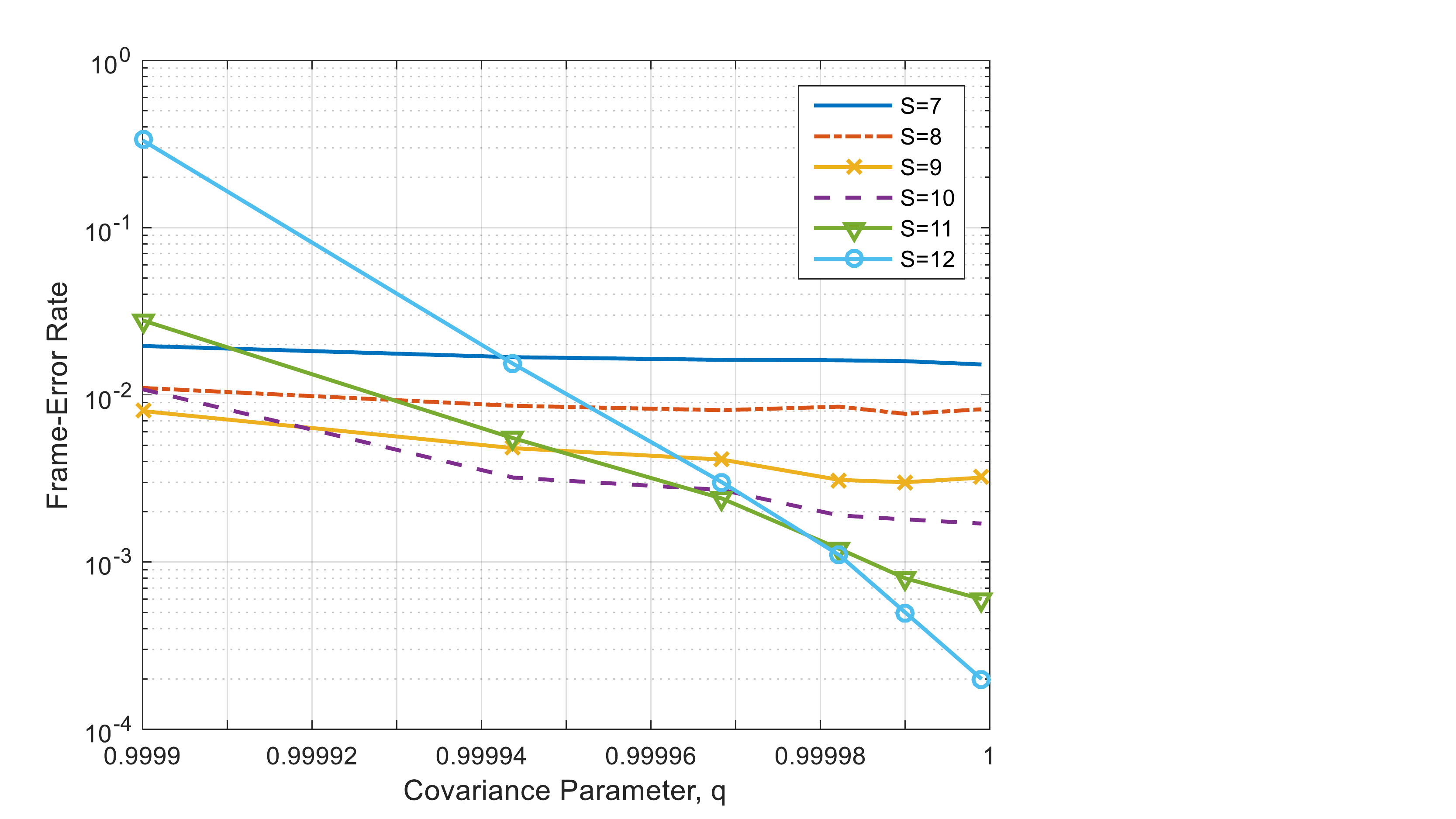}
    \setlength{\belowcaptionskip}{-12pt}
    \caption{FER as a function of $q$ (\mbox{$B \equals 15$.)}}    
    \label{Fig6}
\end{figure}

The FER as a function of the covariance parameter $q$ is plotted in Figs.~\ref{Fig5} and~\ref{Fig6}  for $\mathrm{SNR} \equals 0$ dB  and  payload sizes of $B \equals 5$ bytes and $B \equals 15$ bytes, respectively. The figures demonstrate that for $S \equals 7$ and $S \equals 8$, the performance is not very sensitive to variations in $q$. But for larger spreading factors, which give longer frames for a given given payload size, the value of $q$ has a strong impact on the FER. For example, with $B \equals 15$, $S \equals 11$ provides lower FER than the smaller spreading factors only for  $q \gthan 0.99997$. A comparison of Figs.~\ref{Fig5} and~\ref{Fig6} shows that a larger payload size makes the FER more sensitive to variations in $q$. 

\begin{figure}  
    \centering
    \includegraphics[scale=0.28, bb=-50 30 820 570]{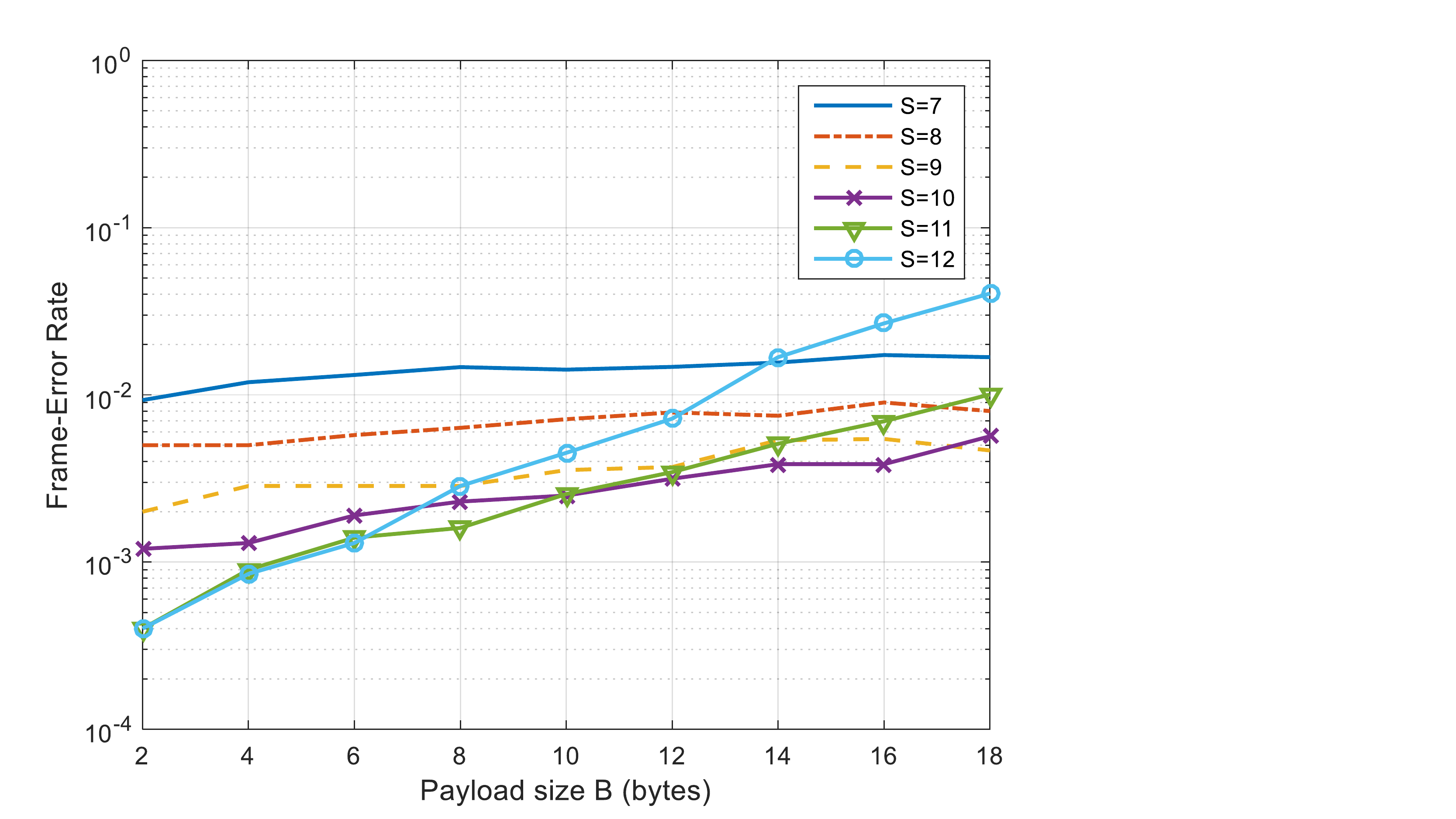}
    \setlength{\belowcaptionskip}{-12pt}
    \caption{FER as a function of $B$ (\mbox{$q \equals 0.99994$}.) }    
    \label{Fig7}
\end{figure}

In Fig.~\ref{Fig7}, we plot the FER for \mbox{$q \equals 0.99994$} as a function of the payload size. This figure confirms our earlier observation that the larger spreading factors are more sensitive to the payload size. For this particular value of $q$, $S \equals 12$ performs worse than all other spreading factors for payload sizes greater than 14 bytes.

\vspace*{-2mm}
\section{Conclusion} \label{conclusion}
In summary, we have observed that the larger spreading factors in LoRa are more susceptible to performance degradations as a result of variations in the channel gain during the reception of a frame. The potential for degradation increases with an increase in the payload size. With certain payload sizes and certain values for the channel's covariance parameter, the performance of spreading factors 11 and 12 can be substantially poorer than the smaller spreading factors. Since the larger spreading factors require more transmission energy owing to their longer durations, they are both energy-inefficient and less reliable in the aforementioned scenarios. Thus, in a fast-fading environment, blindly using a larger spreading factor in the hopes of achieving reliable communications can be counter productive. An optimal spreading-factor-allocation strategy must take into account not only the average SNR but also the fading characteristics and the payload size. 

Future works include investigations of a wider variety of channel models, analytical characterization of the frame-error rate, and the incorporation of interference into the model.

\end{document}